\PassOptionsToPackage{unicode}{hyperref}
\PassOptionsToPackage{hyphens}{url}
\documentclass[
]{article}
\usepackage{amsmath,amssymb}
\usepackage{lmodern}
\usepackage{iftex}
\ifPDFTeX
  \usepackage[T1]{fontenc}
  \usepackage[utf8]{inputenc}
  \usepackage{textcomp} 
\else 
  \usepackage{unicode-math}
  \defaultfontfeatures{Scale=MatchLowercase}
  \defaultfontfeatures[\rmfamily]{Ligatures=TeX,Scale=1}
\fi
\IfFileExists{upquote.sty}{\usepackage{upquote}}{}
\IfFileExists{microtype.sty}{
  \usepackage[]{microtype}
  \UseMicrotypeSet[protrusion]{basicmath} 
}{}
\makeatletter
\@ifundefined{KOMAClassName}{
  \IfFileExists{parskip.sty}{%
    \usepackage{parskip}
  }{
    \setlength{\parindent}{0pt}
    \setlength{\parskip}{6pt plus 2pt minus 1pt}}
}{
  \KOMAoptions{parskip=half}}
\makeatother
\usepackage{xcolor}
\usepackage{longtable,booktabs,array}
\usepackage{calc} 
\usepackage{etoolbox}
\makeatletter
\patchcmd\longtable{\par}{\if@noskipsec\mbox{}\fi\par}{}{}
\makeatother
\IfFileExists{footnotehyper.sty}{\usepackage{footnotehyper}}{\usepackage{footnote}}
\makesavenoteenv{longtable}
\usepackage{graphicx}
\makeatletter
\def\maxwidth{\ifdim\Gin@nat@width>\linewidth\linewidth\else\Gin@nat@width\fi}
\def\maxheight{\ifdim\Gin@nat@height>\textheight\textheight\else\Gin@nat@height\fi}
\makeatother
\setkeys{Gin}{width=\maxwidth,height=\maxheight,keepaspectratio}
\makeatletter
\def\fps@figure{htbp}
\makeatother
\ifLuaTeX
  \usepackage{selnolig}  
\fi
\IfFileExists{bookmark.sty}{\usepackage{bookmark}}{\usepackage{hyperref}}
\IfFileExists{xurl.sty}{\usepackage{xurl}}{} 
\hypersetup{
  hidelinks,
  pdfcreator={LaTeX via pandoc}}

\author{}
\date{}

\begin{document}

\begin{longtable}[]{@{}
  >{\raggedright\arraybackslash}p{(\columnwidth - 0\tabcolsep) * \real{1.0000}}@{}}
\toprule()
\begin{minipage}[b]{\linewidth}\raggedright
\includegraphics[width=3.45972in,height=1.2125in]{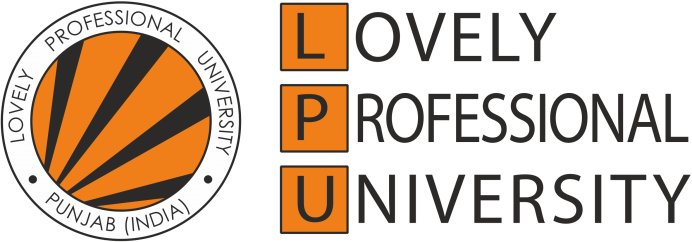}

\textbf{SCHOOL OF POLYTECHNIQUE}

\textbf{2019-2022 Batch}

\textbf{SMART PORTABLE COMPUTER}

\begin{quote}
\textbf{MAJOR PROJECT}\\
Submitted in partial fulfillment of the\\
Requirement for the award of the\\
course of

\textbf{DIPLOMA}\\
\textbf{IN}\\
\textbf{Electronics and Communication Engineering}
\end{quote}

\emph{\textbf{By}}

\textbf{Niladri Das (11918705)}

\begin{quote}
\emph{\textbf{Under the Guidance of}}\\
\textbf{Mr. Ashish Sharma}\\
\textbf{Mrs. Reena Aggarwal}\\
\textbf{Mrs. Meenakshi Gupta}\\
\textbf{Mr. Shagun Sharma}\\
\textbf{Mrs. Navita}
\end{quote}\strut
\end{minipage} \\
\midrule()
\endhead
\bottomrule()
\end{longtable}

\textbf{CERTIFICATE}

This is to certify that the Major project titled ``\textbf{SMART
PORTABLE COMPUTER}'' that is being submitted by ``\emph{Niladri Das}
(11918705)'' is in partial fulfillment of the requirements for the award
of \textbf{COURSE OF DIPLOMA}, is a record of bonafide work done under
my guidance. The contents of this major project, in full or in parts,
have neither been taken from any other source nor have been submitted to
any other Institute or University for award of any degree or diploma and
the same is certified.

The major project is fit for submission and the partial fulfillment of
the conditions for the award of Diploma in Electronics and communication
Engineering.

\textbf{Mr. Ashish Sharma}

\textbf{(Assistant Professor)}

School of Polytechnique\\
Lovely Professional University\\
Phagwara, Punjab.

\textbf{Date: 14-04-2022}

i

\textbf{DECLARATION}

I hereby declare that the project work entitled ``\textbf{SMART PORTABLE
COMPUTER}'' is an authentic record of our own work carried out as
requirements of Major Project for the award of course of Diploma in
Electronics and Communication Engineering from Lovely Professional
University, Phagwara, under the guidance of Assistant Professor Mr.
Ashish Sharma, Mrs. Reena Aggarwal, Mrs. Meenakshi Gupta, Mr. Shagun
Sharma, Mrs. Navita during January-May 2022.

Name: Niladri Das

Registration number: 1191870

ii

\textbf{ACKNOWLEDGEMENT}

I have taken efforts in this project. However, it would not have been
possible without the kind support and help of teachers and the
organizations. I would like to extend our sincere thanks to all of them.
I am highly indebted to Lovely Professional University for their
guidance and constant supervision.

My thanks and appreciation go to my mentor ``Mrs. Meenakshi Gupta'' who
has given me the opportunity for experiencing such knowledge and
providing necessary information regarding the major project and also for
their support in completing this report.

I would like to express my gratitude towards our parents, group mates,
and all the members of the university for their kind cooperation and
encouragement which helped me in completion of this project.

I would like to express my special gratitude and thanks to the
university persons for giving me such attention and time and people who
have willingly helped us out with their abilities. Last but not the
least, we thank ALMIGHTY GOD for completing this project.

Niladri Das (11918705)

iii

\textbf{ABSTRACT}

Due to the covid-19 when most of the organization, school, colleges and
universities have opened as in the virtual platforms in that scenario
many of the students have faced a lot of trouble while purchasing PCs
like Desktop or Laptop because the minimum prices that start with around
25,000 INR and that too consumers would not be able to get a good system
specification on that particular computer and also those who generally
carries their laptop with them for work purposes that is also a
traditional way of carrying a computing device, ``Portable Smart
Computer'' has the speed and performance which is a step up towards the
future of pocket-sized computers. This smart computer provides complete
desktop experience, whether editing documents, browsing the web with a
bunch of tabs open, juggling spreadsheets or drafting a presentation,
we\textquotesingle ll find the experience smooth and very recognizable
but definitely on a smaller, more energy-efficient and much more
cost-effective manner. Programmers will be able to use programming
languages like python, C, C++ and so on and also Keil, Xilinx compilers
could be run easily on this system.

iv

\textbf{TABLE OF CONTENTS}\\
\textbf{Page}
Certificate................................................................................
............i
Declaration................................................
...........................................ii
Acknowledgemaent.....................................
...........................................iii
Abstract....................................................
............................................iv Table of
Contents.......................................
...........................................v List of
Figures...........................................
...........................................vii List of
Tables............................................................................
...........viii

\textbf{CHAPTER}\\
\textbf{1.}
\textbf{INTRODUCTION.......................................................................01}
1.1Motivation
\ldots\ldots\ldots........................................................................................01
1.2Objective of the
study...\ldots\ldots.........................................................................01
1.3Problem Statement
.......................................................................................02

\begin{quote}
\textbf{2.} \textbf{TECHNOLOGIES USED
....................................................................................03}
2.1Raspberry pi 4
...................................................................................................03
2.2Processor
........................................................................................................04
2.3RAM
.......................................................................07\\
2.4Dedicated ROM
...................................................................................08
2.5Wireless
LAN...................................................................................................09
2.6Bluetooth...................................................................................................10
\textbf{3. HARDWARE AND SOFTWARE
DESCRIPTION\ldots\ldots\ldots...............................10}
\end{quote}

v

\begin{quote}
3.1 Block Diagram of the System
...................................................................10
3.2Hardware Requirements for the System
\ldots...................................................10
3.2.1Raspberry Pi 4B
...............................................................11
3.2.2LCD Display
..................................................................14
3.2.3Power Supply
..........................................................................19
3.3Software Requirements for the System
.........................................................20 3.3.1Arduino
IDE
............................................................................20
3.3.2MIT Inventor 2
........................................................................21
3.3.3ThingSpeak
..............................................................................22

3.4Circuit Diagram
..............................................................................................25
3.4.1Diagram
...................................................................................25
3.4.2Explanation of the circuit
.........................................................26
\end{quote}

\textbf{4METHODOLOGY AND
WORKING............................................................................27}
4.1Expected Outcomes of the Project
.................................................................27
4.2Flow Chart
......................................................................................................28
4.3Software description
.......................................................................................29
4.4Working of the System
...................................................................................34

\textbf{5RESULT AND CONCLUSION
\ldots................................................................................36}
5.1Result
...............................................................................................................36
5.2Applications
.....................................................................................................36
5.3Limitations
.......................................................................................................36
5.4Conclusion Future Scope
.................................................................................36

\textbf{6REFERENCES
...............................................................................................................38}

vi

\textbf{List of Figures}

\begin{longtable}[]{@{}
  >{\raggedright\arraybackslash}p{(\columnwidth - 4\tabcolsep) * \real{0.3333}}
  >{\raggedright\arraybackslash}p{(\columnwidth - 4\tabcolsep) * \real{0.3333}}
  >{\raggedright\arraybackslash}p{(\columnwidth - 4\tabcolsep) * \real{0.3333}}@{}}
\toprule()
\begin{minipage}[b]{\linewidth}\raggedright
\textbf{Figure no.}
\end{minipage} & \begin{minipage}[b]{\linewidth}\raggedright
\begin{quote}
\textbf{Figure Name}
\end{quote}
\end{minipage} & \begin{minipage}[b]{\linewidth}\raggedright
\begin{quote}
\textbf{Page no.}
\end{quote}
\end{minipage} \\
\midrule()
\endhead
\begin{minipage}[t]{\linewidth}\raggedright
\begin{quote}
Figure 2.1
\end{quote}
\end{minipage} & \begin{minipage}[t]{\linewidth}\raggedright
\begin{quote}
Raspberry Pi 4B
\end{quote}
\end{minipage} & \begin{minipage}[t]{\linewidth}\raggedright
\begin{quote}
\textbf{3}
\end{quote}
\end{minipage} \\
\begin{minipage}[t]{\linewidth}\raggedright
\begin{quote}
Figure 2.2
\end{quote}
\end{minipage} & \begin{minipage}[t]{\linewidth}\raggedright
\begin{quote}
LCD Screen
\end{quote}
\end{minipage} & \begin{minipage}[t]{\linewidth}\raggedright
\begin{quote}
\textbf{4}
\end{quote}
\end{minipage} \\
\begin{minipage}[t]{\linewidth}\raggedright
\begin{quote}
Figure 2.3
\end{quote}
\end{minipage} & \begin{minipage}[t]{\linewidth}\raggedright
\begin{quote}
Power Supply
\end{quote}
\end{minipage} & \begin{minipage}[t]{\linewidth}\raggedright
\begin{quote}
\textbf{5}
\end{quote}
\end{minipage} \\
Figure 2.4 & \begin{minipage}[t]{\linewidth}\raggedright
\begin{quote}
Block Diagram
\end{quote}
\end{minipage} & \begin{minipage}[t]{\linewidth}\raggedright
\begin{quote}
\textbf{8}
\end{quote}
\end{minipage} \\
\begin{minipage}[t]{\linewidth}\raggedright
\begin{quote}
Figure 3.1
\end{quote}
\end{minipage} & \begin{minipage}[t]{\linewidth}\raggedright
\begin{quote}
Pin Description of Raspberry Pi 4
\end{quote}
\end{minipage} & \begin{minipage}[t]{\linewidth}\raggedright
\begin{quote}
\textbf{10}
\end{quote}
\end{minipage} \\
\begin{minipage}[t]{\linewidth}\raggedright
\begin{quote}
Figure 3.2
\end{quote}
\end{minipage} & \begin{minipage}[t]{\linewidth}\raggedright
\begin{quote}
Circuit Diagram
\end{quote}
\end{minipage} & \begin{minipage}[t]{\linewidth}\raggedright
\begin{quote}
\textbf{12}
\end{quote}
\end{minipage} \\
\bottomrule()
\end{longtable}

vii

\textbf{CHAPTER 1}

\textbf{INTRODUCTION}

\textbf{1.Introduction}

\begin{quote}
We are familiar with Computers in this era. There are many types of
computers; General purpose Computers, Server Computers and also
Supercomputers which are being used for scientific aspect.

In the category of general-purpose computers Desktop computers and
Laptops are still not small in size comparing to Smart Phones. ``Smart
Portable Computer'' is an electronic computing device which can run
Programs, Games, YouTube and also various Software-Applications. With
its mini smart LCD touch display screen, the device comes within a
length of 9cm. The feeling of accessing this device while holding it on
the palm gives tremendous feeling due to its amazing compact design and
lightweight feature.

This Portable Smart Computer carries Processor for its brain, RAM for
dissipating data more quickly and ROM for storing data and also data
could be fetched directly from the External HDD. The data storage option
is not fixed, the computer user will decide whether how much of the data
storage property will be used or needed.

The Portable Computer has 4 bi-directional I/O ports which will be
allowing the system to be connect with Keyboard, Mouse, External HDD and
other I/O devices.

One Gigabit Ethernet port is attached to it and along with those two
micro-HDMI ports are also placed on the circuit to provide two 4K
resolution video output display on HDMI accessible electronic screen.

While taking care of the audio input and output 3.5 mm jack has been
placed on its motherboard.

Power can be given to this smart computer from 5.1V DC 3.0A AC Adaptor
and also from the Power bank through its USB C type power port.
\end{quote}

1

\textbf{1.1. Motivation}

Through this major project there is an enhancement of futuristic
computer technology, so motive was to create it in a portable form and
many for advanced technological aspects could be performed.

\textbf{1.2. Objectives of the study}

SoC carries all the functions of this computing device which has the
memory and processing unit all together interfaced in an architecture.

\textbf{1.3 Problem Statement}

As we look nowadays, Computers are not so much affordable as compared to
mobile phones, so this Smart Portable Computer will feed each and every
execution which we generally are looking for in our day-to-day life like
programming, watching you-tube and so on.

2

\textbf{CHAPTER 2} \textbf{TECHNOLOGIES USED Introduction}\\
In this chapter, we will see technologies used in the designing of the
``Smart Portable Computer''.

\textbf{2.1 Raspberry Pi 4B}\\
The Raspberry Pi 4 is the powerful development of the extremely
successful credit card-sized computer system. This 3rd-generation
Raspberry comes with a high-performance ARM Cortex A72 4x 1.5 GHz quad
core processor. This upgrade gives Raspberry Pi 4 a lot more performance
in certain applications.

\begin{quote}
\includegraphics[width=5in,height=3.75in]{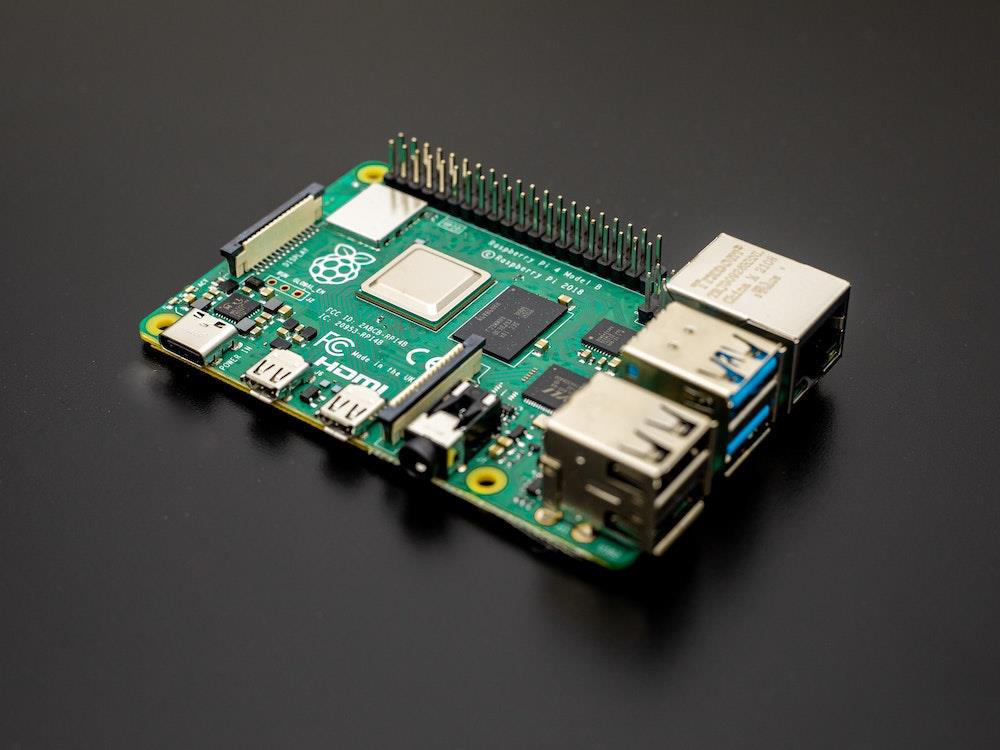}
\end{quote}

Figure 2.1: Raspberry Pi 4B

3

Block Diagram, of the system

\includegraphics[width=6.5in,height=2.76944in]{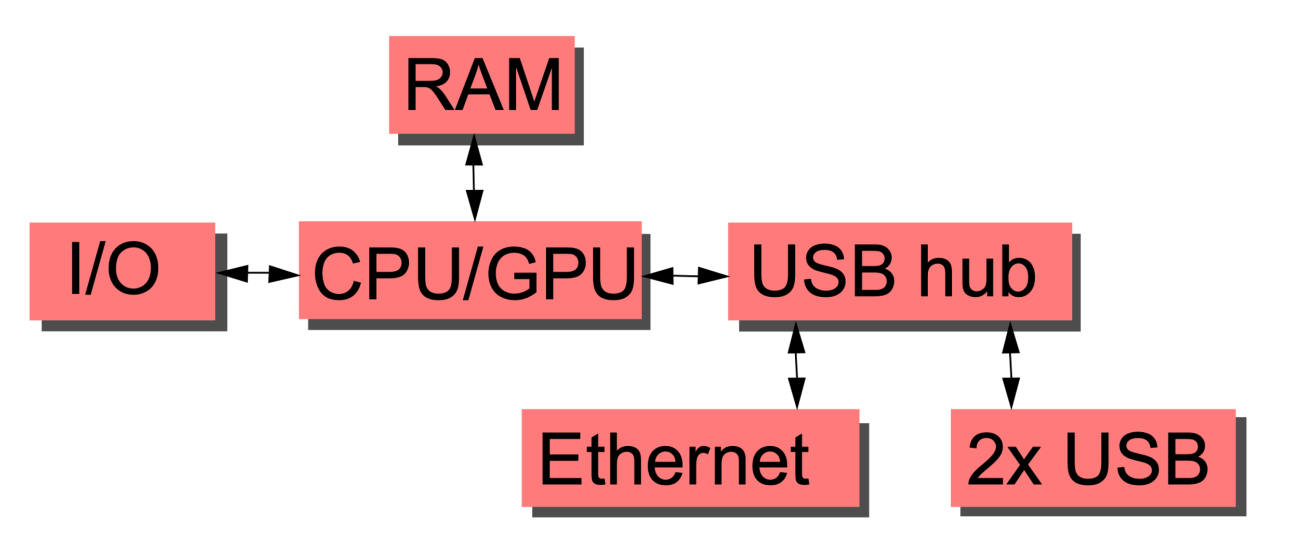}

\begin{quote}
\textbf{2.2 Sensors}\\
We live in a World of Sensors. You can find different types of Sensors
in our homes, offices, cars etc. working to make our lives easier by
turning on the lights by detecting our presence, adjusting the room
temperature, detect smoke or fire, make us delicious coffee, open garage
doors as soon as our car is near the door and many other tasks. All
these and many other automation tasks are possible because of Sensors.
Before going in to the details of What is a Sensor and the Different
Types of Sensors and Applications of these different types of Sensors,
we will first take a look at a simple example of an automated system,
which is possible because of Sensors (and many other components as
well).

\textbf{2.2.1 Real Time Application of Sensors}

The example we are talking about here is the Autopilot System in
aircrafts. Almost all civilian and military aircrafts have the feature
of Automatic Flight Control system or sometimes called as Autopilot.
\end{quote}

4

\includegraphics[width=4.70833in,height=3.02778in]{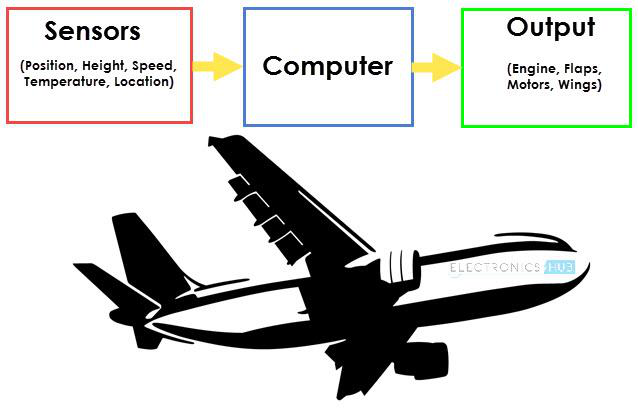}

Figure 2.2: Application of Sensors

\begin{quote}
An Automatic Flight Control System consists of several sensors for
various tasks like speed control, height, position, doors, obstacle,
fuel, maneuvering and many more. A Computer takes data from all these
sensors and processes them by comparing them with pre-designed values.
The computer then provides control signal to different parts like
engines, flaps, and rudders etc. that help in a smooth flight. The
combination of Sensors, Computers and Mechanics makes it possible to run
the plane in Autopilot Mode. All the parameters i.e. the Sensors (which
give inputs to the Computers), the Computers (the brains of the system)
and the mechanics (the outputs of the system like engines and motors)
are equally important in building a successful automated system.

\textbf{2.2.2 Sensor}
\end{quote}

There are numerous definitions as to what a sensor is but I would like
to define a Sensor as an input device which provides an output (signal)
with respect to a specific physical quantity (input).

\begin{quote}
The term ``input device'' in the definition of a Sensor means that it is
part of a bigger system which provides input to a main control system
(like a Processor or a Microcontroller) another unique definition of a
Sensor is as follows: It is a device that converts signals from one
energy domain to electrical domain. The definition of the Sensor can be
understood if we take an example in to consideration.
\end{quote}

5

Figure 2.3 - Types of Sensors

\begin{quote}
The simplest example of a sensor is an LDR or a Light Dependent
Resistor. It is a device; whose resistance varies according to intensity
of light it is subjected to. When the light falling on an LDR is more,
its resistance becomes very less and when the light is less, well, the
resistance of the LDR becomes very high. We can connect this LDR in a
voltage divider (along with other resistor) and check the voltage drop
across the LDR. This voltage can be calibrated to the amount of light
falling on the LDR. Hence, a Light Sensor. Now that we have seen what a
sensor is, we will proceed further with the classification of Sensors.

\textbf{2.2.3 Classification of Sensors}

There are several classifications of sensors made by different authors
and experts. Some are very simple and some are very complex. The
following classification of sensors may already be used by an expert in
the subject but this is a very simple.

In the first classification of the sensors, they are divided in to
Active and Passive. Active Sensors are those which require an external
excitation signal or a power signal. Passive Sensors, on the other hand,
do not require any external power signal and directly generates output
response. The other type of classification is based on the means of
detection used in the sensor. Some of the means of detection are
Electric, Biological, and Chemical, Radioactive etc. The next
classification is based on conversion phenomenon i.e. the input and the
output. Some of the common conversion phenomena are Photoelectric,
Thermoelectric, Electrochemical, Electromagnetic, Thermopolis, etc. The
final classification of the sensors are Analog and Digital Sensors.
Analog Sensors produce an analog output i.e., a continuous output signal
with respect to the quantity being measured. Digital Sensors, in
contrast to Analog Sensors, work with discrete or digital data. The data
in digital sensors, which is used for conversion and transmission, is
digital in nature.

\textbf{2.2.4 Different Types of Sensors}

The following is a list of different types of sensors that are commonly
used in various applications. All these sensors are used for measuring
one of the physical properties like Temperature, Resistance,
Capacitance, Conduction, Heat Transfer etc.

• Temperature Sensor
\end{quote}

6

\begin{longtable}[]{@{}
  >{\raggedright\arraybackslash}p{(\columnwidth - 2\tabcolsep) * \real{0.5000}}
  >{\raggedright\arraybackslash}p{(\columnwidth - 2\tabcolsep) * \real{0.5000}}@{}}
\toprule()
\begin{minipage}[b]{\linewidth}\raggedright
\begin{quote}
•\\
•\\
•\\
•\\
•\\
•\\
•\\
•\\
•\\
•\\
•\\
•
\end{quote}\strut
\end{minipage} & \begin{minipage}[b]{\linewidth}\raggedright
\begin{quote}
Proximity Sensor\\
Accelerometer\\
IR Sensor (Infrared Sensor)\\
Pressure Sensor\\
Light Sensor\\
Ultrasonic Sensor\\
Smoke, Gas and Alcohol Sensor Touch Sensor\\
Color Sensor\\
Humidity Sensor\\
Tilt Sensor\\
Flow and Level Sensor
\end{quote}\strut
\end{minipage} \\
\midrule()
\endhead
\bottomrule()
\end{longtable}

\begin{quote}
\textbf{2.3 Thing Speak Cloud Platform}

ThingSpeak is a platform providing various services exclusively targeted
for building IoT applications. It offers the capabilities of real-time
data collection, visualizing the collected data in the form of charts,
ability to create plugins and apps for collaborating with web services,
social network and other APIs. We will consider each of these features
in detail below.

The core element of ThingSpeak is a `ThingSpeak Channel'. A channel
stores the data that we send to ThingSpeak and comprises of the below
elements:\\
•8 fields for storing data of any type - These can be used to store the
data from a sensor or from an embedded device.

•3 location fields - Can be used to store the latitude, longitude and
the elevation. These are very useful for tracking a moving device.

•1 status field - A short message to describe the data stored in the
channel.
\end{quote}

One of the key elements of an IoT system is an IoT service. ThingSpeak
is one such application platform offering a wide variety of features. At
the heart of ThingSpeak is a channel which can be

used for storing and processing data collected from the `things'.
ThingSpeak also provides various

7

\begin{quote}
apps for integration with web services, other APIs and social networks
and provides the capability to create the applications as plugins. It is
a great platform with extensive possibilities to explore the integration
of the Internet of Things.

\includegraphics[width=5.71528in,height=2.48611in]{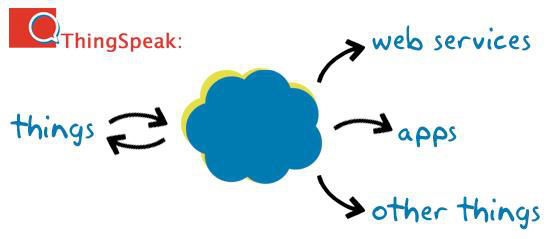}
\end{quote}

Figure 2.4: Overview of ThingSpeak

\begin{quote}
Although the original intended use of the \textbf{ThingSpeak}API is to
\textquotesingle give voice\textquotesingle{} to everyday objects, it
seems that there is a very common emerging trend on what the users are
building and sharing with the IoT. Scholars note that the IoT will be
most useful in an organizational environment, especially for inventory
management, production efficiency, waste management, urban planning,
environmental sensing, social interaction gadgets, continuous care,
emergency response, smart product management, as well as other uses
focusing on creating an efficient and sustainable urban environment. For
individuals and private homes, the IoT is predicted to incorporate smart
metering of electricity, home automation and intelligent shopping
Overall then, the typical applications of any API used to connect
objects to the IoT is broad with far-reaching implications.

\textbf{2.4 Android Application}

Android is an open source and Linux-based operating system for mobile
devices such as smartphones and tablet computers. Android was developed
by the Open Handset Alliance, led by Google, and other companies. This
tutorial will teach you basic Android programming and will also take you
through some advance concepts related to Android application
development. An Android app is a software application running on the
Android platform. Because the Android platform is built for mobile
devices, a typical Android app is designed for a smartphone or a tablet
\end{quote}

8

\begin{quote}
PC running on the Android OS. Although an Android app can be made
available by developers through their websites, most Android apps are
uploaded and published on the Android Market, an online store dedicated
to these applications. The Android Market features both free and priced
apps.

Android apps are written in the Java programming language and use Java
core libraries. They are first compiled to Dalvik executables to run on
the Dalvik virtual machine, which is a virtual machine specially
designed for mobile devices. Novice developers who simply want to play
around with Android programming can make use of the App Inventor. Using
this online application, a user can construct an Android app as if
putting together pieces of a puzzle.
\end{quote}

9

\begin{quote}
\textbf{CHAPTER 3} \textbf{HADWARE AND SOFTWARE DESCRIPTION
Introduction}\\
In this chapter, we are going to discuss the block diagram. And we get
to know the hardware and software requirements needed in the designing
of pet feeder. And we mentioned the circuit diagram which helps us to
understand the design.

\textbf{3.1 Block Diagram}
\end{quote}

\includegraphics[width=6in,height=4.11111in]{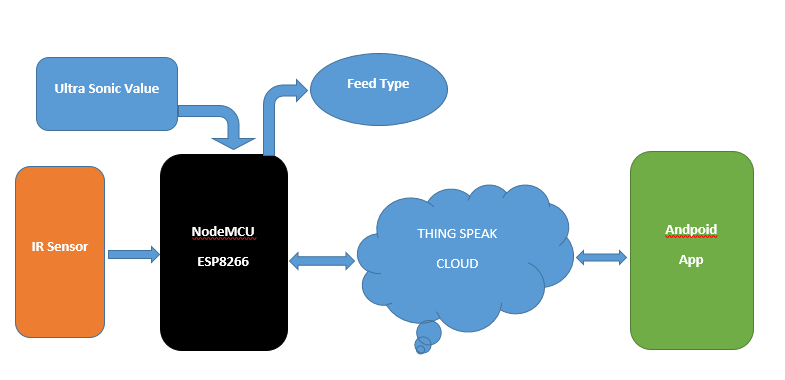}

Figure 3.1 -- Block Diagram\\
Above block diagram we can depict that all Sensors attached to NodeMCU
ESP8266 send the data to it and from there the data will be updated on
ThingSpeak Cloud. And from cloud the information

10

will be read by Android App and response from the application is stored
in the cloud accordingly.

\begin{quote}
From ThingSpeak NodeMCU fetches the date and feed the pet accordingly.

\textbf{3.2 Hardware Requirements of the System}
\end{quote}

Below mentioned are the hardware requirements used in Smart Portable
Computer. And the

\begin{quote}
features and specifications of the hardware components are mentioned.

\textbf{3.2.1 Raspberry Pi 4 B}

Raspberry Pi is a series of small(SBCs) developed in theby thewith The
Rasiginallotion of teaching bin schools and in The original model became
more po\\
anticipide itsfor uses such as It is widely used in many areash as
forause of its low cularity, and open design. It is typically
uselectronic hobbyists, due to its adoption of HDMI and USB devices.

After the release of the second board type, the Raspberry Pi Foundation
set up a new entity, named Raspberry Pi Trading, and installedas with
the responsibility of developing technologyThe Foundatiocatan
educational charity for promoting the teachingasic computer science in
schools and developing countries. Most Pis are made in afactory in
Waleswhile others are made in China and Japan.
\end{quote}

\includegraphics[width=3.71111in,height=2.78194in]{vertopal_9abe5e44f6e2467281f23b493a529b93/media/image2.png}

Figure 3.2: Raspberry Pi 4 B

11

\begin{quote}
\textbf{3.2.1.1 Pin Description}

\textbf{GPIO (General Purpose Input Output) Pins:}\\
NodeMCU has general purpose input output pins on its board as shown in
above pinout diagram. We can make it digital high/low and control things
like LED or switch on it. Also, we can generate PWM signal on these GPIO
pins.

\textbf{ADC (Analog to Digital Converter) channel (A0):} NodeMCU has one
ADC channel/pin on its board.

\textbf{SPI (Serial Peripheral Interface) Pins:}\\
NodeMCU based ESP8266 has Hardware SPI (HSPI) with four pins available
for SPI communication. It also has SPI pins for Quad-SPI communication.
With this SPI interface, we can connect any SPI enabled device with
NodeMCU and make communication possible with it.

\textbf{I2C (Inter-Integrated Circuit) Pins:}\\
NodeMCU has I2C functionality support on ESP8266 GPIO pins. Due to
internal functionality on ESP-12E we cannot use all its GPIOs for I2C
functionality. So, do tests before using any GPIO for I2C applications.

\textbf{UART (Universal Asynchronous Receiver Transmitter) Pins:}\\
NodeMCU based ESP8266 has two UART interfaces, UART0 and UART1. Since
UART0 (RXD0
\end{quote}

\& TXD0) is used to upload firmware/codes to board, we can't use them in
applications while

\begin{quote}
uploading firmware/codes.
\end{quote}

\includegraphics[width=3.08333in,height=2.05556in]{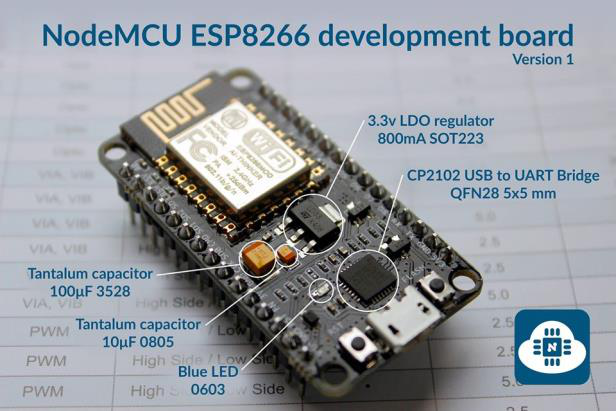}

Figure 3.3 : Development board

12

\begin{quote}
The Development Kit based on ESP8266, integates GPIO, PWM, IIC, 1-Wire
and ADC all in one board.

Power your development in the fastest way combinating with NodeMCU
Firmware!

•USB-TTL included, plug\&play

•10 GPIO, every GPIO can be PWM, I2C, 1-wire •PCB antenna

\includegraphics[width=6.09722in,height=4.5in]{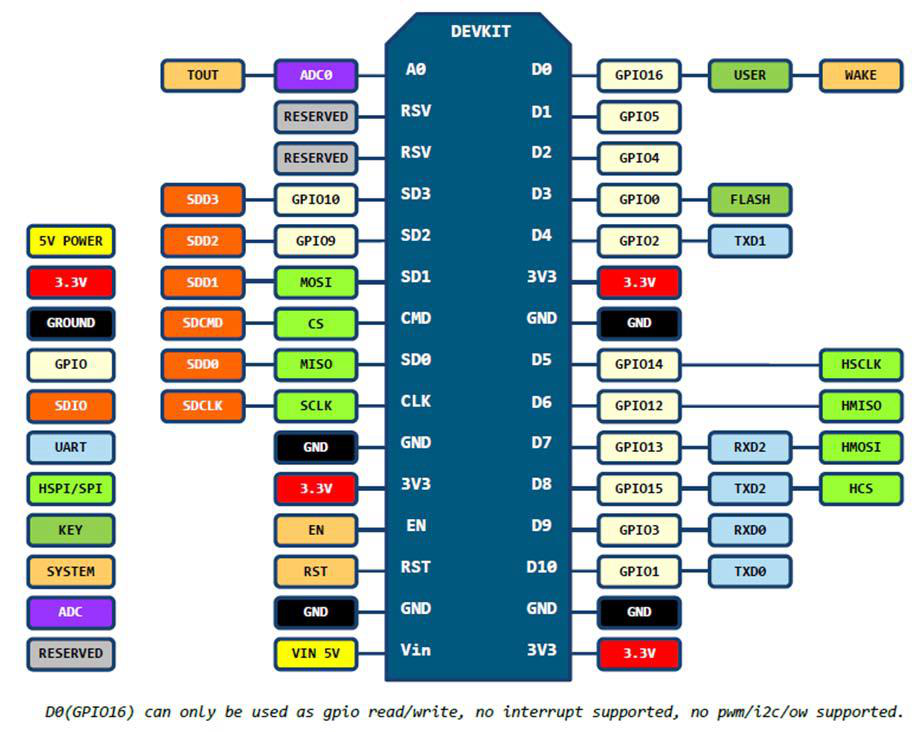}

Figure 3.4 : Pin Description of NodeMCU\\
The most basic way to use the ESP8266 module is to use serial commands,
as the chip is basically a WiFi/Serial transceiver. However, this is not
convenient. What we recommend is using the very cool Arduino ESP8266
project, which is a modified version of the Arduino IDE that you need to
install on your computer. This makes it very convenient to use the
ESP8266 chip as we will be using the well-known Arduino IDE.
\end{quote}

13

\begin{quote}
\textbf{3.2.2 Ultra Sonic Sensor}\\
The HC-SR04 Ultrasonic Sensor is a very affordable proximity/distance
sensor that has been used mainly for object avoidance in various
robotics projects. The HC-SR04 is an ultrasonic ranging module. This
economical measurement functionality with a ranging accuracy that can
reach up to 3mm. Each HC-SR04 module includes an ultrasonic transmitter,
a receiver and a control circuit.

\textbf{3.2.2.1 Ultrasonic Sensor Pin Configuration}
\end{quote}

\begin{longtable}[]{@{}
  >{\raggedright\arraybackslash}p{(\columnwidth - 4\tabcolsep) * \real{0.3333}}
  >{\raggedright\arraybackslash}p{(\columnwidth - 4\tabcolsep) * \real{0.3333}}
  >{\raggedright\arraybackslash}p{(\columnwidth - 4\tabcolsep) * \real{0.3333}}@{}}
\toprule()
\begin{minipage}[b]{\linewidth}\raggedright
\begin{quote}
\textbf{Pin}\\
\textbf{Number}
\end{quote}\strut
\end{minipage} & \begin{minipage}[b]{\linewidth}\raggedright
\begin{quote}
\textbf{Pin Name}
\end{quote}
\end{minipage} & \begin{minipage}[b]{\linewidth}\raggedright
\begin{quote}
\textbf{Description}
\end{quote}
\end{minipage} \\
\midrule()
\endhead
\begin{minipage}[t]{\linewidth}\raggedright
\begin{quote}
1
\end{quote}
\end{minipage} & \begin{minipage}[t]{\linewidth}\raggedright
\begin{quote}
Vcc
\end{quote}
\end{minipage} & \begin{minipage}[t]{\linewidth}\raggedright
\begin{quote}
The Vcc pin powers the sensor, typically with +5V
\end{quote}
\end{minipage} \\
\begin{minipage}[t]{\linewidth}\raggedright
\begin{quote}
2
\end{quote}
\end{minipage} & \begin{minipage}[t]{\linewidth}\raggedright
\begin{quote}
Trigger
\end{quote}
\end{minipage} & Trigger pin is an Input pin. This pin has to be kept
high for 10us to initialize measurement by sending US wave. \\
\begin{minipage}[t]{\linewidth}\raggedright
\begin{quote}
3
\end{quote}
\end{minipage} & \begin{minipage}[t]{\linewidth}\raggedright
\begin{quote}
Echo
\end{quote}
\end{minipage} & \begin{minipage}[t]{\linewidth}\raggedright
\begin{quote}
Echo pin is an Output pin. This pin goes high for a period of time which
will be equal to the time taken for the US wave to return back to the
sensor.
\end{quote}
\end{minipage} \\
\begin{minipage}[t]{\linewidth}\raggedright
\begin{quote}
4
\end{quote}
\end{minipage} & \begin{minipage}[t]{\linewidth}\raggedright
\begin{quote}
Ground
\end{quote}
\end{minipage} & \begin{minipage}[t]{\linewidth}\raggedright
\begin{quote}
This pin is connected to the Ground of the system.
\end{quote}
\end{minipage} \\
\bottomrule()
\end{longtable}

\begin{quote}
Table 1: Pin Configuration of HCSR-04 \textbf{3.2.2.2 HC-SR04 Sensor
Features}
\end{quote}

\begin{longtable}[]{@{}
  >{\raggedright\arraybackslash}p{(\columnwidth - 2\tabcolsep) * \real{0.5000}}
  >{\raggedright\arraybackslash}p{(\columnwidth - 2\tabcolsep) * \real{0.5000}}@{}}
\toprule()
\begin{minipage}[b]{\linewidth}\raggedright
\begin{quote}
•\\
•\\
•\\
•\\
•\\
•\\
•
\end{quote}\strut
\end{minipage} & \begin{minipage}[b]{\linewidth}\raggedright
\begin{quote}
Operating voltage: +5V\\
Theoretical Measuring Distance: 2cm to 450cm Practical Measuring
Distance: 2cm to 80cm Accuracy: 3mm\\
Measuring angle covered: \textless15°\\
Operating Current: \textless15mA\\
Operating Frequency: 40Hz
\end{quote}\strut
\end{minipage} \\
\midrule()
\endhead
\bottomrule()
\end{longtable}

14

It works by sending out a sound wave at ultrasonic frequency and waits
for it to bounce backfrom the object. Then, the time delay between
transmission of sound and receiving of the sound is used to calculate
the distance.

It is done by using the formula:\\
\textbf{Distance} = (speed of sound * Time Delay) / 2\\
We divide the distance formula by 2 because the sound waves travel a
round trip from the sensor and back to the sensor which doubles the
actual distance.

\includegraphics[width=6.26389in,height=2.44444in]{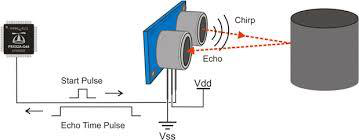}

Figure 3.5: working of ultrasonic sensor.

The Ultrasonic sound waves has an extremely high pitch that humans
cannot hear and is also free from external noises from passive or active
sources. This particular sensor transmits an ultrasonic sound that has a
frequency of about 40 kHz. The sensor has two main parts- transducer
that creates an ultrasonic sound wave while the other part listens to
its echo.

\textbf{3.2.2.2 Applications of Ultrasonic Sensor}: •Loop control.

\begin{quote}
•Roll diameter, tension control, winding and unwind •Liquid level
control\\
•Thru beam detection for high-speed counting\\
•Full detection\\
•Thread or wire break detection
\end{quote}

15

\begin{quote}
\textbf{3.2.3 IR Sensor}

IR Sensor module has great adaptive capability of the ambient light,
having a pair of infrared transmitter and the receiver tube, the
infrared emitting tube to emit a certain frequency, encounters an
obstacle detection direction (reflecting surface), infrared reflected
back to the receiver tube receiving, after a comparator circuit
processing, the green LED lights up, while the signal output will output
digital signal (a low-level signal), through the potentiometer knob to
adjust the detection distance, the effective distance range 2
\textasciitilde{} 10cm working voltage of 3.3V-5V. The detection range
of the sensor can be adjusted by the potentiometer, with little
interference, easy to assemble, easy to use features, can be widely used
robot obstacle avoidance, obstacle avoidance car assembly line count and
black-and-white line tracking and many other occasions.

\textbf{3.2.3.1 Features of IR Sensor Module:}\\
•When the module detects obstacles in front of the signal, the circuit
board green indicator light level, while the OUT port continuous output
low-level signals, the module detects a distance of 2 \textasciitilde{}
10cm, detection angle 35 °, the detection distance can be potential
adjustment with adjustment potentiometer clockwise, the increase in
detection distance; counterclockwise adjustment potentiometer, the
detection distance decreased.

•The sensor module output port OUT can be directly connected with the
microcontroller IO port can also be driven directly to a 5V relay;
Connection: VCC-VCC; GND-GND; OUT-IO.

•The comparator using LM393, stable\\
•3-5V DC power supply module can be used. When the power is turned on,
the red power LED is lit.

•With the screw holes of 3mm, easy to install.
\end{quote}

\includegraphics[width=2.43056in,height=1.66667in]{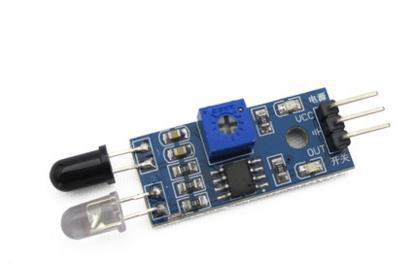}

Figure 3.6 : IR Sensor

16

\begin{quote}
\textbf{Interface(3-wire):-}

•VCC external 3.3V-5V voltage (can be directly connected with the a 5v
microcontroller and 3.3v microcontroller).

•GND external GND.

•OUT board digital output interface (0 and 1).

\textbf{3.2.4 Servo Motor}\\
A servo motor is an electrical device which can push or rotate an object
with great precision. If you want to rotate and object at some specific
angles or distance, then you use servo motor. It is just made up of
simple motor which run through \textbf{servo mechanism}. If motor is
used is DC powered then it is called DC servo motor, and if it is AC
powered motor then it is called AC servo motor. We can get a very high
torque servo motor in a small and light weight packages. Doe to these
features they are being used in many applications like toy car, RC
helicopters and planes, Robotics, Machine etc.
\end{quote}

\includegraphics[width=2.61111in,height=1.86111in]{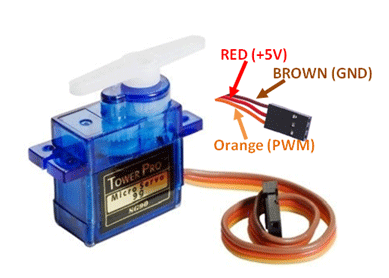}

Figure 3.7 : Servomotor.

\begin{quote}
\textbf{3.2.4.1 Wire Configuration}
\end{quote}

\begin{longtable}[]{@{}
  >{\raggedright\arraybackslash}p{(\columnwidth - 4\tabcolsep) * \real{0.3333}}
  >{\raggedright\arraybackslash}p{(\columnwidth - 4\tabcolsep) * \real{0.3333}}
  >{\raggedright\arraybackslash}p{(\columnwidth - 4\tabcolsep) * \real{0.3333}}@{}}
\toprule()
\begin{minipage}[b]{\linewidth}\raggedright
\begin{quote}
\textbf{Wire}\\
\textbf{Number}
\end{quote}\strut
\end{minipage} & \begin{minipage}[b]{\linewidth}\raggedright
\begin{quote}
\textbf{Wire Colour}
\end{quote}
\end{minipage} & \begin{minipage}[b]{\linewidth}\raggedright
\begin{quote}
\textbf{Description}
\end{quote}
\end{minipage} \\
\midrule()
\endhead
\bottomrule()
\end{longtable}

17

\begin{longtable}[]{@{}
  >{\raggedright\arraybackslash}p{(\columnwidth - 4\tabcolsep) * \real{0.3333}}
  >{\raggedright\arraybackslash}p{(\columnwidth - 4\tabcolsep) * \real{0.3333}}
  >{\raggedright\arraybackslash}p{(\columnwidth - 4\tabcolsep) * \real{0.3333}}@{}}
\toprule()
\begin{minipage}[b]{\linewidth}\raggedright
\begin{quote}
1
\end{quote}
\end{minipage} & \begin{minipage}[b]{\linewidth}\raggedright
\begin{quote}
Brown
\end{quote}
\end{minipage} & \begin{minipage}[b]{\linewidth}\raggedright
\begin{quote}
Ground wire connected to the ground of system
\end{quote}
\end{minipage} \\
\midrule()
\endhead
\begin{minipage}[t]{\linewidth}\raggedright
\begin{quote}
2
\end{quote}
\end{minipage} & \begin{minipage}[t]{\linewidth}\raggedright
\begin{quote}
Red
\end{quote}
\end{minipage} & \begin{minipage}[t]{\linewidth}\raggedright
\begin{quote}
Powers the motor typically +5V is used
\end{quote}
\end{minipage} \\
\begin{minipage}[t]{\linewidth}\raggedright
\begin{quote}
3
\end{quote}
\end{minipage} & \begin{minipage}[t]{\linewidth}\raggedright
\begin{quote}
Orange
\end{quote}
\end{minipage} & \begin{minipage}[t]{\linewidth}\raggedright
\begin{quote}
PWM signal is given in through this wire to drive the motor
\end{quote}
\end{minipage} \\
\bottomrule()
\end{longtable}

\begin{quote}
Table 2: Wire Configuration of Servo motor •TowerPro SG-90 Features\\
•Operating Voltage is +5V typically\\
•Torque: 2.5kg/cm\\
•Operating speed is 0.1s/60°\\
•Gear Type: Plastic\\
•Rotation : 0°-180°\\
•Weight of motor : 9gm\\
•Package includes gear horns and screws

\textbf{3.2.4.2 Applications}\\
•Used as actuators in many robots like Biped Robot, Hexapod, robotic arm
etc.. •Commonly used for steering system in RC toys.

•Robots where position control is required without feedback.

\textbf{3.2.5 Plywood}

Plywood is a sheet material manufactured from thin layers or "plies" of
wood that are glued together with adjacent layers having the Wood grain
rotated up to 90 degrees to one another. It is an from the family of
manufactured boards which includes d(chipboard).

used as a d for the petfeeder circuit. This prevents pets damaging the
system.
\end{quote}

18

\includegraphics[width=2.04167in,height=2.04167in]{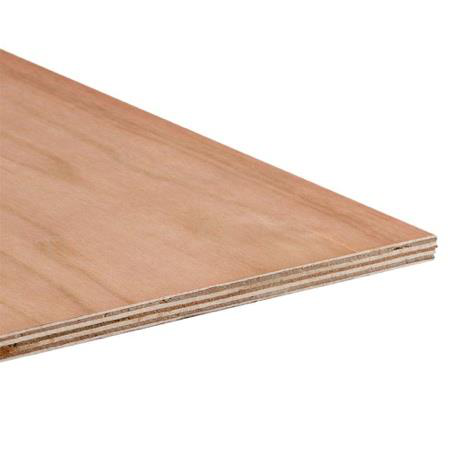}

Figure 3.8 : Plywood

\begin{quote}
\textbf{3.2.6 Power Supply}\\
The NodeMCU ESP8266 can be powered via the USB connection or with an
external power supply. The power source is selected automatically.
External (non-USB) power can come either from an AC-to-DC adapter
(wall-wart) or battery. The adapter can be connected by plugging a 2.1mm
center-positive plug into the board\textquotesingle s power jack. Leads
from a battery can be inserted in the Gnd and Vin pin headers of the
POWER connector.

The board can operate on an external supply of 6 to 20 volts. If
supplied with less than 7V, however, the 5V pin may supply less than
five volts and the board may be unstable. If using more than 12V, the
voltage regulator may overheat and damage the board. The recommended
range is 7 to 12 volts.

The power pins are as follows:

•\textbf{VIN.} The input voltage to the Arduino board when
it\textquotesingle s using an external power source (as opposed to 5
volts from the USB connection or other regulated power source). You can
supply voltage through this pin, or, if supplying voltage via the power
jack, access it through this pin.

•\textbf{3V3.} A 3.3 volt supply generated by the on-board regulator.
Maximum current draw is 50 mA.

•\textbf{GND.} Ground pins.
\end{quote}

19

\begin{quote}
\textbf{3.3 Software Requirements for the System}

\textbf{3.3.1 Arduino IDE}

The Arduino Integrated Development Environment - or Arduino Software
(IDE) - contains a text editor for writing code, a message area, a text
console, a toolbar with buttons for common functions and a series of
menus. It connects to the Arduino and Genuino hardware to upload
programs and communicate with them.

The Arduino(IDE) is aapplication (for ogramming la originated fronand It
includes a code edito features such as text cutting and pasting, se
reptext, automatic indenting, and and provides simple one-click
mechanisms to com pro board. It also contains a message area, a text
console, a toolbar with buttons for common functions and a hierarchy of
operation menus. The source code for the IDE is released under the
version 2.

The Arduino IDE using special rules of code structuring. The Arduino IDE
supplies afm thproject, which provides many common input and output
procedn code onires two basic functions, for starting the sketch and the
main program loop, that are compiled and linked with a program stub
main() into an executableprogram with the also included with the IDE
distribution. Tmploys the progranvert the executable code into a text
file in hexadecimal encoding that is loaded into the Arduino board by a
loader program in the board\textquotesingle s firmware.
\end{quote}

20

\includegraphics[width=3.69444in,height=3.26389in]{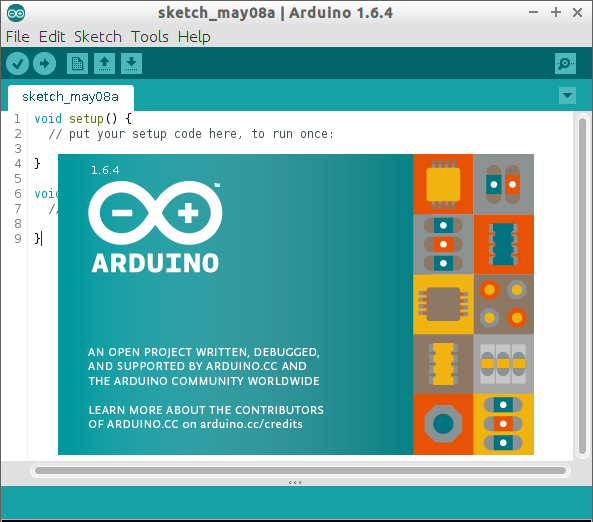}

Figure 3.9: Arduino IDE Interface

\begin{quote}
\textbf{3.3.2 MIT Inventor}

App Inventor lets you develop applications for Android phones using a
web browser and either a connected phone or emulator. The App Inventor
servers store your work and help you keep track of your projects.

MIT App Inventor is an intuitive, visual programming environment that
allows everyone -- even children -- to build fully functional apps for
smartphones and tablets. Those new to MIT App Inventor can have a simple
first app up and running in less than 30 minutes. And
what\textquotesingle s more, our blocks-based tool facilitates the
creation of complex, high-impact apps in significantly less time than
traditional programming environments. The MIT App Inventor project seeks
to democratize software development by empowering all people, especially
young people, to move from technology consumption to technology
creation.

The App Inventor development environment is supported for Mac OS X,
GNU/Linux, and Windows operating systems, and several popular Android
phone models. Applications created with App Inventor can be installed on
any Android phone.
\end{quote}

21

\begin{longtable}[]{@{}
  >{\raggedright\arraybackslash}p{(\columnwidth - 2\tabcolsep) * \real{0.5000}}
  >{\raggedright\arraybackslash}p{(\columnwidth - 2\tabcolsep) * \real{0.5000}}@{}}
\toprule()
\begin{minipage}[b]{\linewidth}\raggedright
\begin{quote}
\includegraphics[width=2.48611in,height=3.73611in]{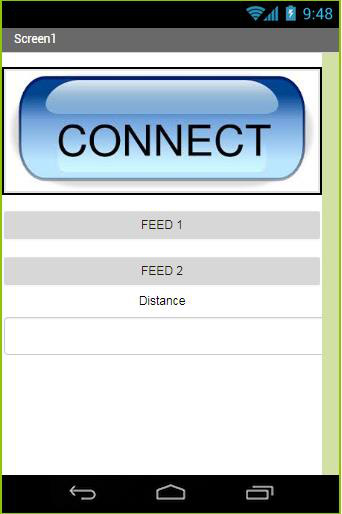}
\end{quote}
\end{minipage} & \begin{minipage}[b]{\linewidth}\raggedright
\begin{quote}
\includegraphics[width=2.48611in,height=3.47222in]{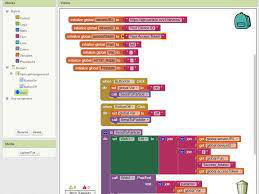}
\end{quote}
\end{minipage} \\
\midrule()
\endhead
\bottomrule()
\end{longtable}

Figure 3.10: MIT Inventor App \& Code blocks

\begin{quote}
\textbf{3.3.3 ThingSpeak}

The term ``Internet of Things'' (IoT), coined by KevinAshton in 1999,
has been in use for several years and continues to be of interest,
specifically when it comes to technological progress. But what exactly
is the IoT? Essentially, it refers to giving objects representation in
thedigital realm through giving them a unique ID and connecting them in
a network. In other words, these things are connected to the internet
and are able to automatically transfer data without relying on human
interaction. Hence being ``Machine to Machine'' (M2M) interaction.
Essentially, M2M interaction enables networked devices to exchange data
and perform actions without the input or assistance of humans, for
instance in remote monitoring. Indeed, the lack of necessity for human
intervention seems to conjure dystopian images ofthe future. But this is
not necessarily the case. For instance,

one can envision the IoT to become an important feature of the `home of
the future', where one can begin pre-heating the oven just before they
get home from work via a (mobile) application. Or perhaps, automatically
turning on the washing machine when the power grid has less load, as
communicated by a remote power plant. Or, businesses can anticipate when
a popular item is
\end{quote}

22

\begin{quote}
running low on stock due to notification from the shelves that they sit
on. Hence, the IoT has many interesting applications that can be applied
to both individuals and corporations.

In order to connect an object to the IoT, several things are needed in
the hardware and software realm. First of all, if one wishes to go
beyond simply connecting data from a computer, objects to gather
(sensors) or receive (actuators) data are necessary. For example, a
digital thermometer can be used to measure temperature. In this case,
the data needs to be uploaded to a network of connected servers which
run applications. Such a network is commonly referred to as `the cloud'.
The cloud utilizes the process of visualization, meaning that several
physical servers can be connected and used in tandem, but appear to the
user as one machine (despite that at the physical level, the machines
function independently). This method of computing thus allows changes to
be

made to the `virtual' server (such as software updates or changes in
storage space) much easier than before.

For the purpose of connecting an object to the IoT, we focus on the
\emph{ThingSpeak} API. In this project we have created two channels
which helps android application and NodeMCU to communicate. This
channels provide the data logs for information retrieval.

\includegraphics[width=5.30556in,height=3.16667in]{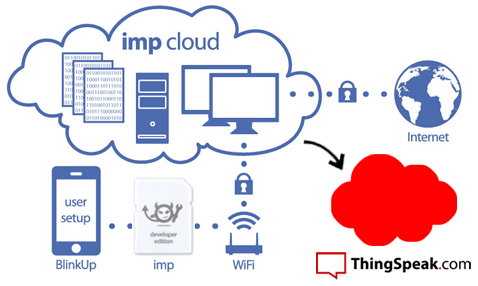}
\end{quote}

Figure 3.11 : ThingSpeak.com

23

\begin{quote}
\textbf{3.4 Circuit Diagram}\\
\textbf{3.4.1 Diagram}
\end{quote}

24

\includegraphics[width=6.90278in,height=3.86111in]{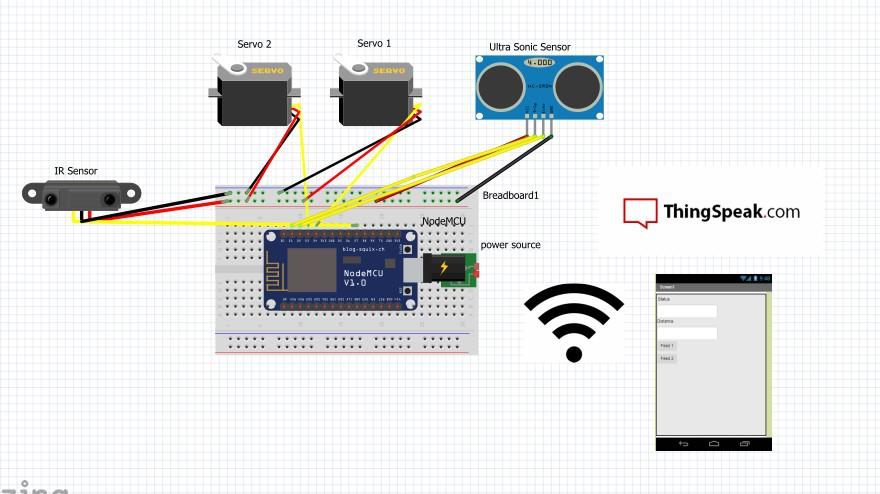}

Figure 3.12 -- Circuit Diagram

\begin{quote}
\textbf{Wiring}

In the above circuit diagram the colors of wires represents the
following usage:\\
Red wire - This color wire represents the 5V power to the circuit from
the NodeMCU. Black wire - This color wire represents the GND provided to
the circuit from NodeMCU.

Yellow wire - This color wire represents the connections between the
output pins of sensors, servo motors and to digital pins of NodeMCU.

\textbf{3.4.2Explanation}

We choose to do this project with NodeMCU because it is an Open Sorce
Platform for IOT which is able to connect to any cloud platform through
its integrated ESP8266 Wi-Fi module. And this is a less cost device
which used less power for processing. The sensors and motors attached to
the pins of NodeMCU in the following manner:

•D1 pin of NodeMCU is connected to the ECHO pin of the ultra-sonic
sensor and D2 pin is connected to TRIG pin of the sensor.
\end{quote}

25

\begin{quote}
•D3 and D4 pins of the NodeMCU are connected to the two different output
pins of Servo motors.

•D7 pin of the NodeMCU is connected to output pin of the IR Sensor.

Through the programmed network NodeMCU get connected to the ThingSpeak
website which monitors the system through the android application and
responds accordingly.
\end{quote}

\textbf{CHAPTER 4}

26

\textbf{METHODOLOGY AND WORKING}

\begin{quote}
\textbf{Introduction}

In this chapter, we are going to discuss the outcomes we got from the
design and we have also mentioned the flowchart involved in the working
of the product. And we have mentioned usage of individual softwares used
in the design.

\textbf{4.1 Expected Outcomes of the System}

•IOT based pet feeder make use of existing technologies like sensors,
motors, android applications and cloud based platforms which helps to
automate the feeding process of pets. With these technologies combined
one can control the system from anywhere from home. This system help pet
owner to take care of their pets at anytime from anywhere. •IOT based
pet feeder enables the process of feeding of the pets easy with less
cost and with known technology it helps everyone to operate easily.

•As in the today market usage of automated pet feeders is increasing but
his model is cost effective compared to the products available in the
market.

•The android application created for this system helps us to know about
the quantity of food available in the bowl and with that we can feed the
pet accordingly.

•In this we have use multi-feed availability for the pet which can be
choose by the pet's owner.

\textbf{4.2 Flow chart}
\end{quote}

27

\includegraphics[width=3.29167in,height=5.61111in]{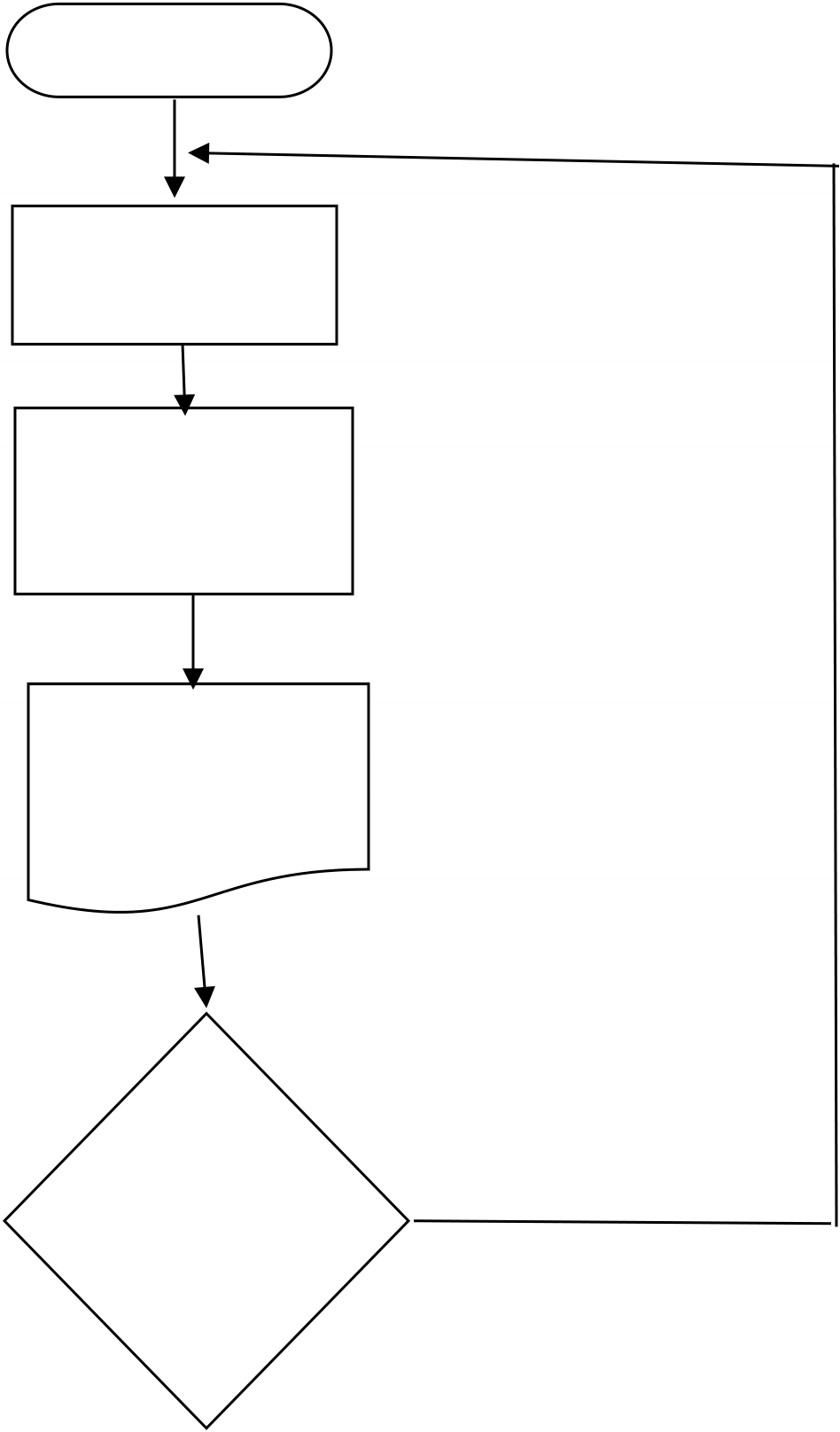}\includegraphics[width=1.375in,height=2.31944in]{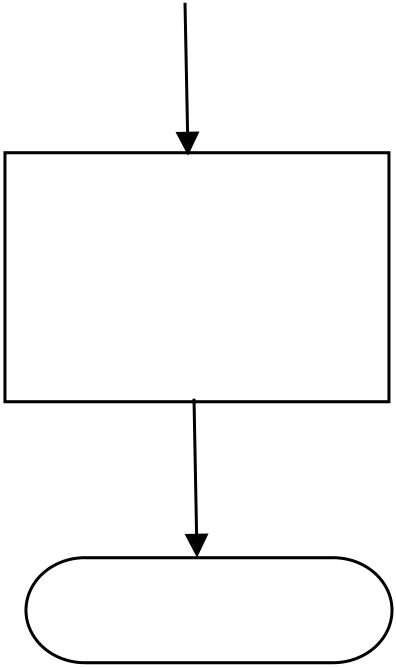}

Start

\begin{quote}
Detection of\\
Pet

Notifies the\\
owner through\\
App

Decides the type\\
of feed from\\
App

If filled if filled Bowl\\
alread\\
y\\
filled
\end{quote}

If not filled

\begin{quote}
Fills the Bowl\\
until it get filled
\end{quote}

End

28

\textbf{4.3 Software description}

In this IOT based pet feeder main technology used is ThingSpeak which
monitors the system based on the logic prgrammed on NodeMCU and helps to
communicate with android application on smart phone. In following
sections we will be discussing the Arduino IDE source code, MIT Inventor
blocks and the channels created on ThingSpeak.

\textbf{Arduino IDE Source Code}

\textbf{MIT Inventor 2}\\
We have used MIT Inventor 2 for designing of the android application
because it is an easily understandable interface for developing an
application.

In MIT Inventor, We use blocks for developing of the applications. It is
a simple drag and drop interface for designing of the application. The
code blocks and designer layout used for the design of application looks
as follows:

Figure 4.1: Code Blocks of App

\textbf{ThingSpeak}\\
In this project we choose ThingSpeak as a cloud platform because it is
an open source IOT platform where we can uptade our data and retrieve
them whenever needed. This is a free of cost platform which helps us to
connect our microcontroller or NodeMCU to internet for automation
purpose.

ThingSpeak provides channels, where we can store our data in the form of
fields. Which are represented graphically and can be viewed publically
or privately. It provides API keys for updating and fetching of the
data. In this project we have used two channels one for updating the

29

data into the cloud which is having two fields and other for getting
information from android application which is used to control the
system.

But the problem associated with tThingSpeak is, It takes 15 seconds to
update the server because of which we see the delay of communication
between android application and NodeMCU. This can be overcome by renting
a commercial account in ThingSpeak.

Below are the channels created in the ThingSpeak:

\begin{quote}
\includegraphics[width=6.18056in,height=2.15278in]{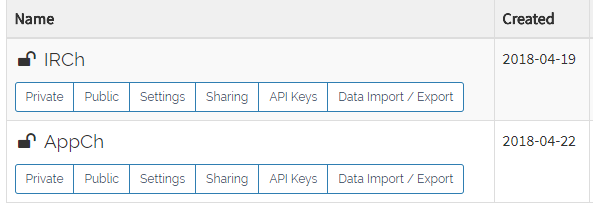}
\end{quote}

Figure 4.2: Channels in ThingSpeak

Below are the graphical representation of fields of 1st channel i.e; IR
Channel

\includegraphics[width=6.5in,height=2.20833in]{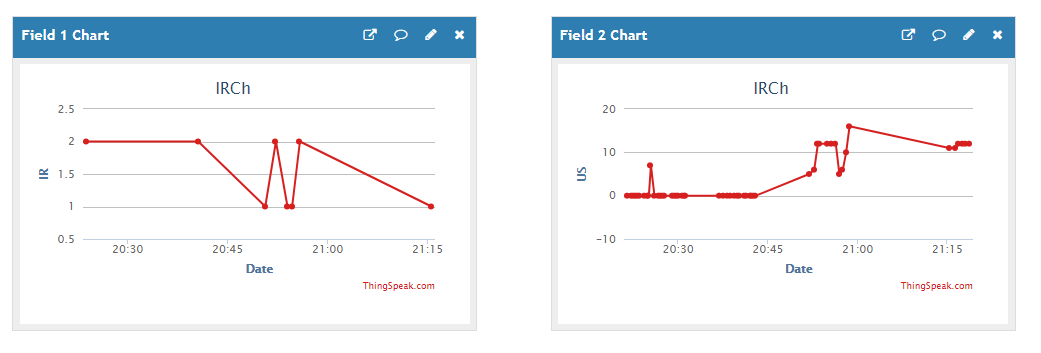}

Figure 4.3: Field Graphs of IRCh in ThingSpeak

Below is the graphical representation of field in he 2nd channel i.e;
AppChannel

30

\begin{quote}
\includegraphics[width=5.41667in,height=1.91667in]{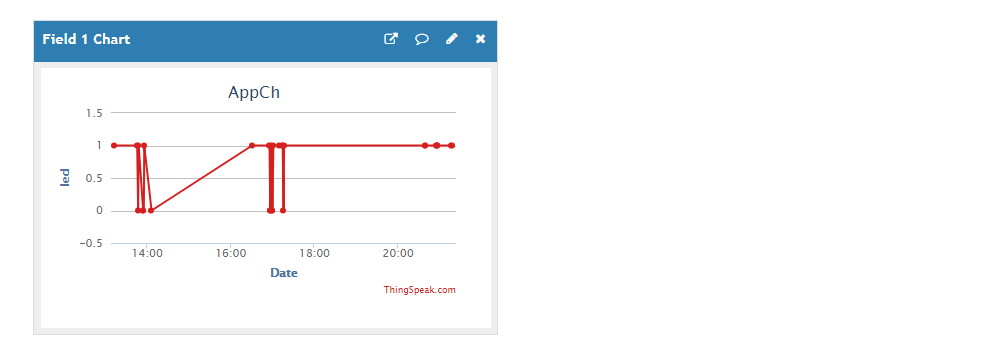}
\end{quote}

Figure 4.4: Field Graphs of IRCh in ThingSpeak

\textbf{4.4 Working of the System}

IOT based pet feeder is designed in the way that it feeds the pet
automatically by the pet's owner from anywhere using an android
application. We have used Ultra Sonic sensor, IR sensor, Servo motor and
NodeMCU to acheieve the design. The whole system is covered by a 2x2
feet wooden box. Below are steps involved in working of the system.

First using IR sensor, we detect the presence of the pet and updates the
value to ThingSpeak IR Channel by using NodeMCU which connects to the
internet and from there we can read the status of the pet in android
application.

\includegraphics[width=1.45833in,height=2.58333in]{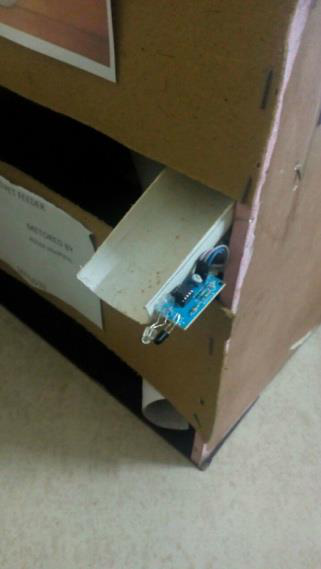}

Figure 4.5: IR Sensor attached to pet feeder

31

And according to the feed decided by the owner get updated in AppChannel
and from their NodeMCU request the action performed from the android
application and responds accordingly.

If owner choose to give feed1 the corresponding servo motor to which
feed is attached get rotated and feed get dropped into the bowl. And
same happens when owner chooses the other feed.

\includegraphics[width=2.31944in,height=2.86111in]{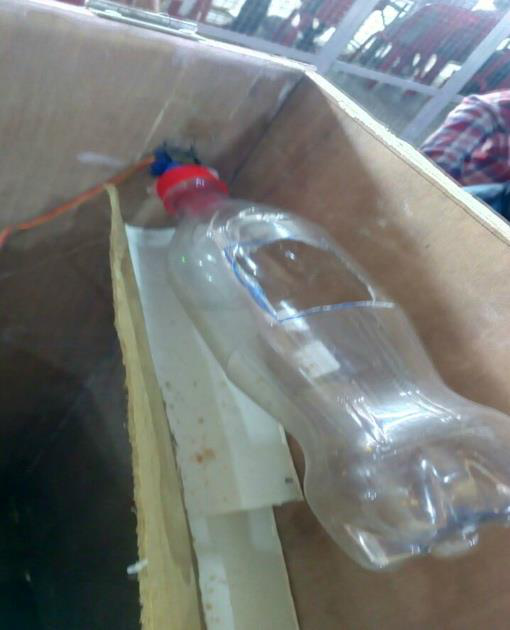}

Figure 4.6: bottle attached to servo

The servo motor rotates until the bowl having feed gets filled fully,
this will be measured by the Ultra Sonic Sensor attached above the bowl.
Once the bowl got filled servo automatically stops rotating.

\includegraphics[width=2.68056in,height=2.38889in]{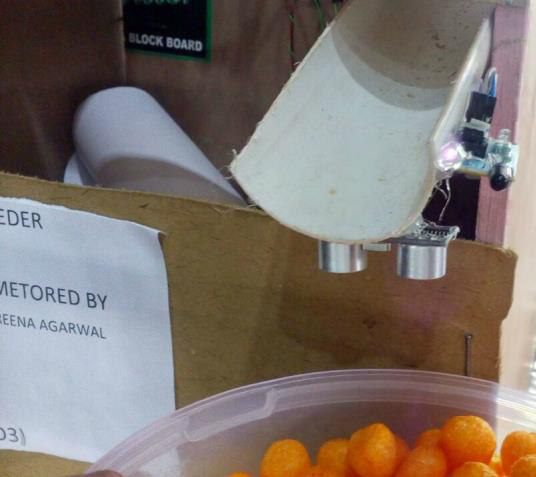}

\begin{quote}
Figure 4.7: Ultra Sonic attached to pet feeder\\
\textbf{CHAPTER 5}
\end{quote}

32

\textbf{RESULT AND CONCLUSION Introduction}\\
In this chapter, we are going to conclude the results we have achieved
through the design. And discussed the limitations and future scope of
deign.

\textbf{5.1 Result}

With this IOT based pet feeder, It is easy for pet owner's to take of
the pet's even if they are not present at home. And to make this cost
effective we have used sensors which are available anywhere. As a
result, It also helps pets to get habitual to automated feeding. The
concrete results of this design are as follows:\\
•Less cost\\
•Easy maintainance\\
•Low power consumption

\textbf{5.2 Applications}\\
•IOT based pet feeder can be used in homes where there are more than one
pet.

\begin{quote}
•And helps people who want to take care of their pets from their work.

•It can be used pet care industries to feed lots of pets simultaneously.

•It keeps track of data of feeding intervals of pet which helps to know
the health status of pet.
\end{quote}

\textbf{5.3 Limitations}\\
As we are using ThingSpeak as cloud platform, this provide a 15 seconds
delay to respond to the operation performed from android application. In
this project we are not using any video streaming feature. So, It is
complicated to know the delivery of feed. But this can be overcomed by
renting a commercial server from ThingSpeak.

\textbf{5.4 Conclusion and Future Scope}

33

The interaction between humans and physical devices and objects is
attracting increasing attention. Many studies have attempted to provide
a natural and intuitive approach to request services. The current trend
of combining pet control and IoT technology offers exciting future
developments. The proposed system is also referred on smart-home
technology, including the smart pet feeder. The results not only present
the key improvement of the IOT based pet feeder involved in the IoT
technology, but also meet the demand of pet owners. The basic vision
behind the IoT, it may have a new way of operational method, it may have
a new method of connecting devices, and there might be the even complete
clean-slate approach. As the full operational definition is not yet
finalized, there are numerous research issues that can be worked on. As
a next step, we will fully integrate the other pet care devices into our
system, including litter boxes, pet cam, etc. With that, the diverse
needs of the owners can be met, and the health, monitor, and
entertainment topics for pets are all covered. Besides, standing as the
cloud term, how to connect the numerous networking devices around the
globe is the next issue. In the future, we have to centralize on the
study of the IoT gateway and long-distance detection of the pets.

\textbf{\textbackslash{}}

\textbf{CHAPTER 6}

34

\textbf{REFERENCES}

35

\end{document}